\documentclass[aps,twocolumn,showpacs,amsmath,pra]{revtex4}
\usepackage{hyperref} 
\usepackage{amsmath}
\usepackage{amssymb}
\usepackage{graphicx}
\newcommand{\ket}[1]{\ensuremath{\left|{#1}\right\rangle}}

\newcommand{\bsy}[1]{\ensuremath{\boldsymbol{#1}}}
\newcommand{\brm}[1]{\ensuremath{\mathbf{#1}}}

\newcommand{\sinc}{\ensuremath{\mathrm{sinc}}}

\newcommand{\ue}[1]{\ensuremath{\boldsymbol{\hat{\epsilon}}}_{#1}}

\begin{document} 
\title{Transverse spatial and frequency properties of two-photon 
states generated by spontaneous parametric
down-conversion} 
\author{A. G. da Costa Moura}
\affiliation{Departamento de F\'{\i}sica, Universidade Federal de 
Minas Gerais, Caixa Postal 702, Belo Horizonte, MG 30123-970, Brazil}
\author{W. A. T. Nogueira}
\affiliation{Departamento de F\'{\i}sica, Universidade Federal de 
Minas Gerais, Caixa Postal 702, Belo Horizonte, MG 30123-970, Brazil}
\author{C. H. Monken}
\affiliation{Departamento de F\'{\i}sica, Universidade Federal de 
Minas Gerais, Caixa Postal 702, Belo Horizonte, MG 30123-970, Brazil}
\author{S. P. Walborn}
\affiliation{Instituto de F\' \i sica, Universidade Federal do Rio de
Janeiro, Caixa Postal 68528, Rio de Janeiro, RJ 21945-970, Brazil}
\date{\today}

\begin{abstract}
We present a detailed account of the two-photon states
generated by SPDC in both type I and type II phase matching, including
the effects of anisotropy of the nonlinear medium and the frequency
spread of the down-converted fields.   Accurate as well as simplified
expressions are derived for type I and type II phase matching in the
context of Fourier Optics. The main results are compared with 
experimental data available in the literature, showing good 
agreement in all cases.
\end{abstract}
\maketitle
\section{Introduction}
In the course of the last three decades, spontaneous parametric
down-conversion (SPDC) has proven to be a valuable tool in the
experimental investigation of fundamental properties of the
electromagnetic field in the quantum domain 
\cite{mandel95,shih95,zeilinger},
including nonclassical correlations, entanglement and nonlocality.
The fact that SPDC is capable of generating pairs of photons in a wide
range of frequencies and wave vectors, in addition to the fact that
these photons may be entangled in a number of different degrees of
freedom, qualifies SPDC also as a unique tool in the demonstration of
quantum information procedures and protocols \cite{white}.

In many applications, especially in the pioneering ones, a basic
knowledge of a few properties of the two-photon states generated by
SPDC is enough to explain the effects discussed.  As the applications
become more sophisticated, for example in those that combine
correlations in more than one degree of freedom, a more comprehensive
approach is necessary in order to enhance the capabilities of quantum
state engineering.  In this direction, the use of Fourier Optics
concepts in two-photon optics represents a significant step.

An important application field of SPDC is the construction of
entangled photon state sources with high fidelity and purity
\cite{kwiat95,kwiat99,kurt01,bitton02,shih03,vanexter,altep05,barbieri05}.
In order to obtain efficient and dependable two-photon sources, a
number of questions must be addressed, as for example, phase
compensation, mode coupling, pair collection efficiency, maximum
photon flux, etc.  To be optimized, all these points require a
detailed knowledge of the SPDC process.

Another context in which a more precise model is crucial refers to the
issue of conservation of orbital angular momentum (OAM) of light in
the process of SPDC 
\cite{arnaut00,arnold02,barbosa02,walborn04,torres05,barbosa07,osorio08}.
Oversimplifications or inconsistent assumptions about the two-photon
state generated have led to conflicting conclusions on whether OAM is
conserved or not in SPDC.

In this work, we present a detailed account of the two-photon states
generated by SPDC in both type I and type II phase matching, including
the effects of anisotropy of the nonlinear medium and the frequency
spread of the down-converted fields.  As an exact approach to this
problem presents a considerable level of difficulty, we are forced to
restrict ourselves to some approximations.  We adhere to the commonly
adopted simplified procedure for the quantization of the
electromagnetic field in the nonlinear medium by just multiplying the
$\brm{k}$ vectors by the corresponding refractive indices and keeping
the vacuum expressions for the field operators.  The frequency spread
of the down-converted fields is considered to be small compared to
their central frequencies .  We also work in the paraxial
approximation, in which the propagation of the fields is much easier
to deal with.  Fortunately, these approximations, are good enough to
encompass a great number of practical cases.  We admit, however, that
while the model discussed here is helpful in far field applications,
it may not be appropriate in discussions of the fundamental
interactions inside the nonlinear medium.  Whenever possible, the
results reported here are compared with experimental data, obtained in
our labs or reported in the literature.

It should be mentioned that other authors have reported their
contributions to this field, in different levels of detail and scope
\cite{tang,klyshko,hm,rubin,joobeur,arnaut00,%
kurtsiefer01,torres04,barbosa07,fedorov08,baek08}.
From our point of view, a comprehensive description of SPDC that
accounts for all features observed so far in two-photon states, with
expressions ready to use in Fourier optics, is still missing.  It is
in this context that we present our work.

\section{Two-photon state generated by SPDC}
The interaction Hamiltonian describing the optical processes of
parametric down-conversion in nonlinear birefringent crystals
is usually written in terms of a simplified field quantization in
matter \cite{mandelwolf}.  The field modes participating in this
process are coupled by the second-order susceptibility tensor
$\chi^{(2)}$.  Due to the phase matching conditions, the coupling occurs
only for some sets of polarizations, known as type I and type II. In
negative uniaxial crystals ($n_{e}<n_{o}$) such as BBO
($\beta-$BaB$_{2}$O$_{4}$) and Lithium Iodate (LiIO$_{3}$), the
down-conversion process with type I phase matching can be summarized
as $e\rightarrow oo$, meaning that one photon of the pump beam with
extraordinary polarization is converted into two photons with ordinary
polarization.  In type II phase matching, the process is represented
by $e\rightarrow oe$ or $e\rightarrow eo$.

In a perturbative approach, the (post-selected) two-photon state
generated by SPDC is written as \cite{hm}
\begin{equation}
\label{hong}
\ket{\Psi}=\sum_{\sigma_{1},\sigma_{2}}
\sum_{\brm{k}_{1},\brm{k}_{2}}
\Phi_{\brm{k}_{p}\brm{k}_{1}\brm{k}_{2}
\sigma_{p}\sigma_{1}\sigma_{2}}
\ket{\brm{k}_{1},\sigma_{1}}\ket{\brm{k}_{2},\sigma_{2}},
\end{equation}
where $\brm{k}_{1}$ and $\brm{k}_{2}$ are the wave vectors of the
down-converted fields, $\brm{k}_{p}$ is the wave vector of the pump
field, whose frequency $\omega_{p}$ is assumed to be well defined.
$\sigma_{j}$ indicates the polarization of each down-converted field,
that can be ordinary ($o$) or extraordinary ($e$), and
$\ket{\brm{k}_{j},\sigma_{j}}$ represent a one-photon state in the
plane wave mode $\brm{k}_{j},\sigma_{j}$.  Considering that the
nonlinear crystal is a rectangular block of sides $L_{x}$, $L_{y}$,
$L_{z}$, with two faces parallel to the plane $xy$, the amplitude
$\Phi$ is given by
\begin{eqnarray}
\label{ampli1}
\Phi_{\{\brm{k},\sigma\}} & = &  g_{\brm{k}_{p},\sigma_{p}}\
g^{*}_{\brm{k}_{1},\sigma_{1}}\ g^{*}_{\brm{k}_{2},\sigma_{2}}\ \tau\
e^{i\Omega(t-\frac{\tau}{2})}\ \sinc \frac{\Omega\, \tau}{2} \nonumber
\\
& & \times\ \sum_{i,j,k=x,y,z}\
\tilde{\chi}_{ijk}^{(2)}(\ue{\brm{{k}}_0,\sigma_0})_i
(\ue{\brm{{k}}_1,\sigma_1})_{j}^{*}
(\ue{\brm{{k}}_2,\sigma_2})_{k}^{*} \nonumber \\
& &\times\ \mathcal{E}_{\brm{k}_{p},\sigma_{p}}\
\int_{\mathcal{I}}e^{-i\bsy{\Delta}\cdot \brm{r}}\ d\brm{r},
\end{eqnarray}
where $\{\brm{k},\sigma\}$ represents the set of indices
$\brm{k}_{p}\brm{k}_{1}\brm{k}_{2} \sigma_{p}\sigma_{1}\sigma_{2}$,
$g_{\brm{{k}}_j,\sigma_j} = i \left[\frac{\hbar
\omega_{j}}{2\varepsilon_0 V
n^2(\brm{k}_{j},\sigma_{j})}\right]^{\frac{1}{2}}$, $V$ is the
quantization volume, $\mathcal{I}$ is the interaction volume
($L_{x}L_{y}L_{z}$), $n(\brm{k}_{j},\sigma_{j})$ is the refractive
index corresponding to the mode $\brm{k}_{j},\sigma_{j}$, $\Omega =
\omega_{1}+\omega_{2}-\omega_{p}$, $\tau$ is the interaction time,
$\tilde{\chi}_{ijk}^{(2)}$ is related to the second-order nonlinear
susceptibility tensor \cite{mandelwolf},
$(\ue{\brm{{k}}_i,\sigma_i})_{j}$ ($j=x,y,z$) are the cartesian
components of the polarization vectors,
$\mathcal{E}_{\brm{k}_{p},\sigma_{p}}$ is the pump field amplitude in
the mode $\brm{k}_{p},\sigma_{p}$, and $\bsy{\Delta} =
\brm{k}_{1}+\brm{k}_{2}-\brm{k}_{p}$.

In order to simplify expression (\ref{ampli1}), it is convenient to
make the following approximations:

(\textit{a}) The pump frequency $\omega_{p}$ is well defined and the
interaction time is long, so that the term $\sinc\, \Omega\,
\tau/2$ is significant only when $\omega_{1}+\omega_{2}=\omega_{p}$.
The frequencies $\omega_{1}$ and $\omega_{2}$ can therefore be 
written as
\begin{subequations}
\begin{eqnarray}
\omega_{1}&=&\frac{\omega_{p}}{2}\, (1+\nu),\\
\omega_{2}&=&\frac{\omega_{p}}{2}\, (1-\nu).
\end{eqnarray}
\end{subequations}
This assumption can be justified by the use of a moderate power
continuous-wave pump laser so that the time interval between two
down-conversions is large compared to the detection resolving time.

(\textit{b}) The frequency spread of the detectable down-converted
fields is small compared to the central frequency ( $|\nu|\ll 1$), so
that the dispersion of the refractive indices around the central
frequency $\omega_{p}/2$ is small and a linear approximation can be
used.  This assumption is justified by the use of narrow-band
interference filters in front of the detectors.  For example, for a
100nm wavelength spread centered at 700nm, $\nu$ lies in the
interval ($-0.08,0.08$).

(\textit{c}) The terms $g_{\brm{{k}}_j,\sigma_j}$ and
$\tilde{\chi}_{ijk}^{(2)}$ are slowly-varying functions of
$\brm{k}_{j}$, so that they may be taken as constants in the intervals 
considered for $\brm{k}_{j}$.

(\textit{d}) The pump beam propagates along the $z$ axis and the
crystal is large enough in the $x$ and $y$ directions to contain the
whole pump beam transverse profile.  In this case, $L_{x}$ and $L_{y}$
can be extended to infinity and the last term in expression
(\ref{ampli1}), the integral $\int_{\mathcal{I}}e^{-i\bsy{\Delta}\cdot
\brm{r}}\ d\brm{r}$, is proportional to
\begin{displaymath}
\delta(k_{1x}+k_{2x}-k_{px})\ \delta(k_{1y}+k_{2y}-k_{py}) 
\int_{z_{c}-L_{z}/2}^{z_{c}+L_{z}/2}e^{-i\Delta_{z}z}\ dz,
\end{displaymath}
where $z_{c}$ locates the center of the crystal.

(\textit{e}) The quantization volume is large enough to justify the
replacement of summations in $\brm{k}$ by integrals.  

(\textit{f}) The pump beam contains only extraordinary polarization.
It is implicit in this assumption that we are dealing with negative
birefringent crystals.  

Under the above assumptions, Eq.  (\ref{hong}) is written as
\begin{eqnarray}
\label{psi2}
\ket{\Psi}&=&\sum_{\sigma_{1},\sigma_{2}}\int\!d\nu \int\!d\brm{q}_{1}
\int\!d\brm{q}_{2}\ 
\Phi_{\sigma_{1}\sigma_{2}}(\brm{q}_{1},\brm{q}_{2},\nu)\nonumber \\
&&\times \ket{\brm{q}_{1},\nu,\sigma_{1}}\ket{\brm{q}_{2},-\nu,
\sigma_{2}},
\end{eqnarray}
where $\ket{\brm{q}_{j},\nu,\sigma_{j}}$ represents a one-photon state
in the mode defined by the transverse ($xy$) component $\brm{q}_{j}$
of the wave vector, the frequency $(1+\nu)\omega_{p}/2$ and by the
polarization $\sigma_{j}$.  The amplitude $\Phi$ is now reduced to
\begin{equation}
\label{ampli2}
\Phi_{\sigma_{1}\sigma_{2}}\approx C_{\sigma_{1}\sigma_{2}}\ G(\nu)\ 
\tilde{\mathcal{E}}(\brm{q}_{1}+\brm{q}_{2})\
\int_{z_{c}-L_{z}/2}^{z_{c}+L_{z}/2}e^{-i\Delta_{z}z}\ dz,
\end{equation}
where $C_{\sigma_{1}\sigma_{2}}$ is a coupling constant, which depends
on the nonlinear susceptibility tensor, $G(\nu)$ is the spectral
function defined by the narrow bandwidth filters placed in front of
the detectors, and $\tilde{\mathcal{E}}$ is the plane wave spectrum of
the pump beam $\tilde{\mathcal{E}}(\brm{q}_{p})$, with $\brm{q}_{p}$
replaced by $\brm{q}_{1}+ \brm{q}_{2}$.  The integral in $z$ is
discussed in what follows.  The function $G(\nu)$ may be centered at
zero in frequency degenerate configurations or at some fixed small
detuning $\nu_{0}$.  

Most of the properties of the down-converted
fields is determined by the longitudinal wave vector mismatch
\begin{equation}
\Delta_{z}=k_{1z}+k_{2z}-k_{pz}.
\end{equation}
If the anisotropy of the medium is neglected (which is not always
convenient), $\nu=0$ (monochromatic approximation), and the crystal is
cut for collinear phase matching, $\Delta_{z}$ has a simple expression
in terms of $\brm{q}_{1}$ and $\brm{q}_{2}$, with the $z$ components
replaced by $k_{jz}=\sqrt{|\brm{k}_{j}|^{2}-|\brm{q}_{j}|^{2}}$.  In
the paraxial approximation, $k_{jz}$ can be approximated by
$k_{jz}\approx |\brm{k}_{j}|-|\brm{q}_{j}|^{2}/2|\brm{k}_{j}|$.  This
leads, for $z_{c}=0$, to
\begin{displaymath}
\int_{-L_{z}/2}^{+L_{z}/2}e^{-i\Delta_{z}z}\ dz \propto \sinc
\left(\frac{L_{z}}{4|\brm{k}_{p}|}|\brm{q}_{1}-\brm{q}_{2}|^{2}\right),
\end{displaymath}
which appears in the amplitude reported in Ref.  \cite{monken98}.
This result, however, is a good approximation only when the crystal is
very thin, the diffraction (Rayleigh) length $z_{o}$ of the pump beam
is large, and the detection area is small so that the sinc function
can be approximated by 1.  In more general conditions, the anisotropy
of the crystal must be taken into account.  

\begin{figure}[hbt]
\begin{center}
\includegraphics{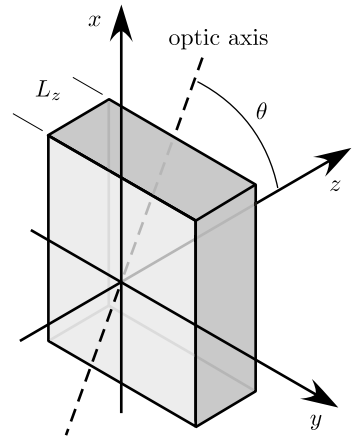}
\caption{\label{fig:slab}Geometry of the problem.  The uniaxial medium
is assumed to be a slab of thickness $L_{z}$, having its faces
parallel to the $xy$ plane.  The optic axis lies in the $xz$ plane, at
an angle $\theta$ with the $z$ direction.}
\end{center}
\end{figure}

From this point on, we will refer to the arrangement depicted in
Fig.\ref{fig:slab}.
\section{The effect of anisotropy}

Let us consider a monochromatic electromagnetic field in a plane wave
mode $e^{i(\brm{k}\cdot\brm{r}-\omega t)}$ propagating through a
nonmagnetic uniaxially birefringent medium with ordinary and
extraordinary refractive indices $n_{o}$ and $n_{e}$, respectively.
Considering that the optic axis lies on the plane $xz$, making an
angle $\theta$ with the $z$ direction, and that the plane wave has
extraordinary polarization, Maxwell's equations require that the
components of the wave vector $\mathbf{k}$, whose cartesian components
are ($q_{x},q_{y},k_{z}$), satisfies the \textit{ray surface}
equation \cite{bornwolf}

\begin{align}\label{eq:elipsoiderodada}
&(n_{o}^{2}\cos^{2}\theta+n_{e}^{2}\sin^{2}\theta)q_{x}^{2}+
(n_{o}^{2}\sin^{2}\theta+n_{e}^{2}\cos^{2}\theta)k_{z}^{2}+
\nonumber \\
&2(n_{o}^{2}-n_{e}^{2})\sin\theta\cos\theta\, q_{x}\, k_{z}
+n_{o}^{2}q_{y}^{2}= \left(n_{o}n_{e}\frac{\omega}{c}\right)^{2}.
\end{align}
Solving Eq. (\ref{eq:elipsoiderodada}) for $k_{z}$, we find
\begin{equation}
\label{eq:kz}
k_{z}=-\alpha q_{x}+\sqrt{\varkappa^{2}-\beta q_{x}^{2}-\gamma q_{y}^{2}},
\end{equation}
where
\begin{equation}
\alpha=\frac{(n_{o}^{2}-n_{e}^{2})\sin\theta 
\cos\theta}{n_{o}^{2}\sin^{2}\theta+n_{e}^{2}
\cos^{2}\theta},
\end{equation}
\begin{equation}
\beta=\left(\frac{n_{o}n_{e}}{n_{o}^{2}\sin^{2}\theta+
n_{e}^{2}\cos^{2}\theta}\right)^{2},
\end{equation}
\begin{equation}
\gamma=\frac{n_{o}^{2}}{n_{o}^{2}\sin^{2}\theta+n_{e}^{2}\cos^{2}\theta},
\end{equation}
\begin{equation}
\varkappa= \eta\frac{\omega}{c},
\end{equation}
with
\begin{equation}
\label{eq:k1}
{\eta=\frac{n_{o}n_{e}}{\sqrt{n_{o}^{2}\sin^{2}\theta+
n_{e}^{2}\cos^{2}\theta}}}.
\end{equation}
The ordinary and extraordinary refractive indices $n_{o}$ and $n_{e}$ 
are obtained from the Sellmeier equations. For BBO, $n_{o}$ and 
$n_{e}$ are obtained from \cite{niko}
\begin{subequations}
\begin{eqnarray}
n_{o}^{2}&=&2.7359 + \frac{0.01878}{\lambda^2 - 0.01822} - 0.01354 
\,\lambda^2,\\
n_{e}^{2}&=&2.3753 + \frac{0.01224}{\lambda^2 - 0.01667} - 0.01516 
\,\lambda^2,
\end{eqnarray}
\end{subequations}
with $\lambda$ given in $\mu$m.
For Lithium Iodate we use \cite{sellmeier}
\begin{subequations}
\begin{eqnarray}
n_{o}^{2}&=&2.083648 + \frac{1.332068\,\lambda^{2}}{\lambda^2 -
0.035306} - 0.008525 \,\lambda^2,\qquad \\
n_{e}^{2}&=&1.673463 + \frac{1.245229\,\lambda^{2}}{\lambda^2 -
0.028224} - 0.003641 \,\lambda^2.
\end{eqnarray}
\end{subequations}
\par
\begin{figure}
\centerline{\includegraphics{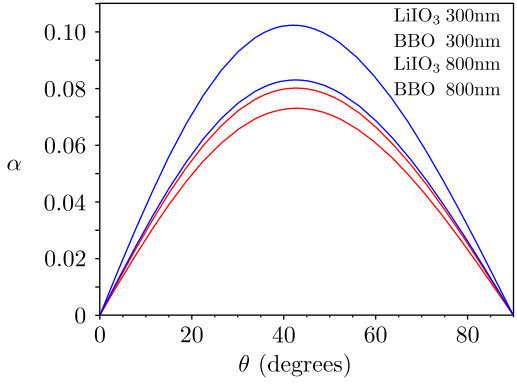}}
\caption{\label{alpha}(Color online) Values of $\alpha$ for for 300nm
and 600nm in BBO and Lithium Iodate as functions of the phase matching 
angle $\theta$. The curves are ordered from top to bottom, according 
to the legend.}
\end{figure}
\begin{figure}
\centerline{\includegraphics{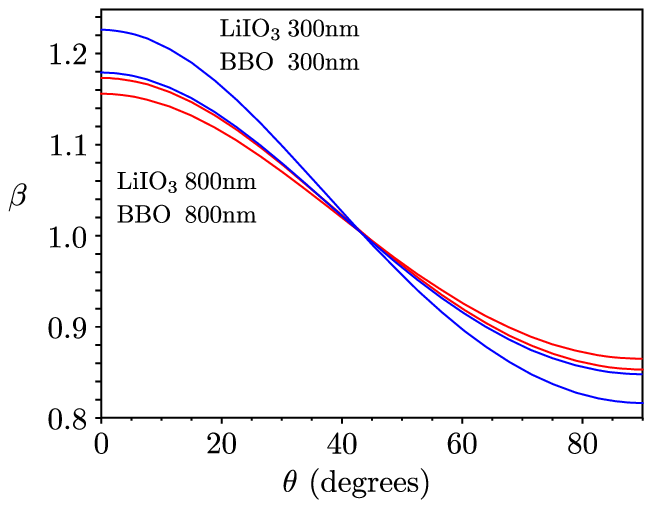}}
\caption{\label{beta}(Color online) Values of $\beta$ for for 300nm
and 600nm in BBO and Lithium Iodate as functions of the phase matching 
angle $\theta$. The curves are ordered from top to bottom, according 
to the legend.}
\end{figure}
\begin{figure}
\centerline{\includegraphics{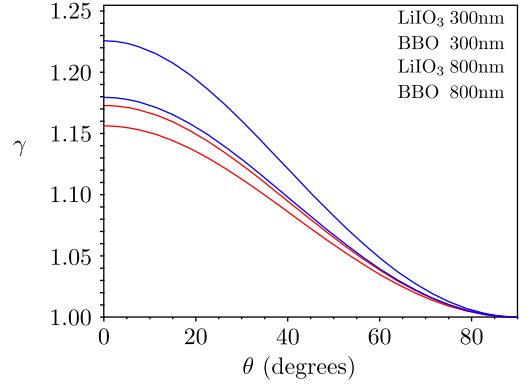}}
\caption{\label{gamma}(Color online) Values of $\gamma$ for for 300nm
and 600nm in BBO and Lithium Iodate as functions of the phase matching 
angle $\theta$. The curves are ordered from top to bottom, according 
to the legend.}
\end{figure}

Within the paraxial approximation,
\begin{equation}
\label{eq:aprox_paraxial_kz}
k_{z}\approx \varkappa-\alpha q_{x}-\frac{1}{2\varkappa}(\beta
q_{x}^{2}+\gamma q_{y}^{2}).
\end{equation}
The term $\alpha$ represents a linear displacement in the $x$
direction, also known as the \textit{walk-off}.  It reaches its
maximum value in the neighborhood of $\theta=45^{\circ}$.  The terms
$\beta$ and $\gamma$ account for the deviation of the curvatures of the
ray surface from a spherical surface in the $x$ and $y$ directions, 
respectively. Differently from $\alpha$, 
the terms $\beta$ and $\gamma$ have little influence in the phase 
matching. The term $\eta$ is the refractive index for a plane wave 
with extraordinary polarization, propagating along the $z$ direction, 
whose wavenumber is $\varkappa$. Figs. \ref{alpha}, \ref{beta} 
and \ref{gamma} show the values of $\alpha$, $\beta$ and $\gamma$ as 
functions of the phase matching angle $\theta$ for 300nm and 600nm in 
BBO and Lithium Iodate.
\section{Type I phase matching}
Let us consider first the case of type I phase matching, where the
pump field has extraordinary polarization and the down-converted
fields have ordinary polarization ($e\rightarrow oo$). In this case,
\begin{equation}
k_{pz}\approx \varkappa_{p}-\alpha_{p}
q_{px}-\frac{1}{2\varkappa_{p}}(\beta_{p}
q_{px}^{2}+\gamma_{p} q_{py}^{2}),
\label{k0z}
\end{equation}
\begin{equation}
k_{1z}\approx k_{1}-\frac{q_{1}^{2}}{2k_{1}},
\end{equation}
\begin{equation}
k_{2z}\approx k_{2}-\frac{q_{2}^{2}}{2k_{2}},
\end{equation}
where $k_{1}=n_{o1}\,\omega_{1}/c$ and $k_{2}=n_{o2}\,\omega_{2}/c$.
Since we are assuming $\nu\ll 1$, the refractive indices $n_{o1}$ and
$n_{o2}$ can be written in a linear approximation as
\begin{subequations}
\begin{eqnarray}
n_{o1}&=&\bar{n}_{o}(1+a\nu),\\
n_{o2}&=&\bar{n}_{o}(1-a\nu),
\end{eqnarray}
\end{subequations} 
where $\bar{n}_{o}$ is the ordinary refractive index at the
frequency $\omega_{p}/2$, and factor $a$ is given by
\begin{equation}
\label{defa}
a=\frac{\omega_{p}}{2\bar{n}_{o}}\left.\frac{dn_{o}}{d\omega}\right 
|_{\omega=\omega_{p}/2}.
\end{equation}

In order to calculate $\Delta_{z}$, which in type I phase matching will 
be referred to as $\Delta^{oo}_{z}$, we have to
write $k_{1z}+k_{2z}-k_{pz}$ replacing $q_{px}$ and $q_{py}$ in Eq.
(\ref{k0z}) by $q_{1x}+q_{2x}$ and $q_{1y}+q_{2y}$, respectively. 
After some manipulation, we find
\begin{eqnarray}
\label{fI}
\Delta^{oo}_{z}&\approx&
K\mu_{oo}+ \alpha_{p}
({q}_{1x}+{q}_{2x})\nonumber\\
&& -\frac{1}{\bar{n}_{o}K}\Big [ \frac{q_{1}^{2}}{1+\nu} +
\frac{q_{2}^{2}}{1-\nu}\nonumber\\
& &-\frac{b}{2}(q_{1x}+q_{2x})^{2}-\ \frac{g}{2}(q_{1y} +
q_{2y})^{2}\Big ],\nonumber\\
\end{eqnarray}
where 
$K=\omega_{p}/c$,
$b=\beta_{p}\bar{n}_{o}/\eta_{p}$, and
$g=\gamma_{p}\bar{n}_{o}/\eta_{p}$.  The subscript $p$ means that
$\alpha$, $\beta$ and $\gamma$ refer to the pump field. $\mu_{oo}$  
is the type I \textit{collinear index mismatch},  defined as
\begin{equation}
\mu_{oo}=\bar{n}_{o}(1+a\nu^{2})-\eta_{p}.
\end{equation}
In the derivation of Eq.  (\ref{fI}), terms in $\nu^{2}$ were
neglected with respect to $\nu$.  For type I collinear ($\mu_{oo}=0$)
phase matching in BBO pumped with $\lambda_{p}=351$nm,
$\alpha_{p}=0.0747$, $a=0.02$, $b=1.06$, and $g=1.11$.  For Lithium
Iodate, $\alpha_{p}=0.0871$, $a=0.03$, $b=0.951$ and $g=1.07$.  

\begin{figure}
\centerline{\includegraphics{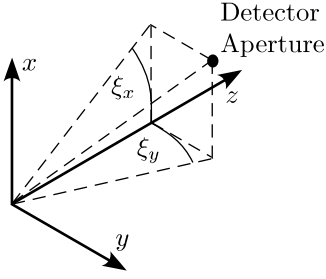}}
\caption{\label{xi}Angular components of the down-converted fields.}
\end{figure}

Sometimes, instead of the transverse components of the wave vectors,
one is interested in the output angles of the down-converted photons. 
In general, the output fields of interest propagate at small angles 
with respect to the $z$ axis, so that one can approximate the output 
angles by
\begin{subequations}
\label{xiq}
\begin{eqnarray}
\xi_{1j}&\approx& \frac{c}{\omega_{1}}q_{1j}= 
\frac{2\,q_{1j}}{(1+\nu)K},\\
\xi_{2j}&\approx& \frac{c}{\omega_{2}}q_{2j}= 
\frac{2\,q_{2j}}{(1-\nu)K},
\end{eqnarray}
\end{subequations}
where $j=x,y$, and we define the vectors
$\bsy{\xi}_{1}=(\xi_{1x},\xi_{1y})$ and
$\bsy{\xi}_{2}=(\xi_{2x},\xi_{2y})$ as shown in Fig. \ref{xi}.
In the far field, $\bsy{\xi}_{1}$ and $\bsy{\xi}_{2}$ can be written 
in terms of the  transverse coordinates $\bsy{\rho}_{1}$ and 
$\bsy{\rho}_{2}$ of the detectors as
\begin{subequations}
\label{xirho}
\begin{eqnarray}
\bsy{\xi}_{1}& \approx & \frac{1}{z_{D}}\bsy{\rho}_{1},\\
\bsy{\xi}_{2}& \approx & \frac{1}{z_{D}}\bsy{\rho}_{2},
\end{eqnarray}
\end{subequations}
where $z_{D}$ is the $z$ coordinate of the detection plane (parallel
to the $xy$ plane). 

In terms of the angular variables $\bsy{\xi}_{1}$ and 
$\bsy{\xi}_{2}$, the amplitude $\Phi_{oo}$, defined in Eq. 
(\ref{ampli2}), is written as
\begin{eqnarray}
\label{phiang}
\Phi_{oo}& \approx & N\,G(\nu)\,\tilde{\mathcal{E}}\left[\frac{K}{2}(\bsy{\xi}_{1} +
\bsy{\xi}_{2})+\frac{K\nu}{2}(\bsy{\xi}_{1} - \bsy{\xi}_{2})\right]\nonumber\\
&& \times \sinc\left[\frac{KL_{z}}{2}\
f_{oo}(\bsy{\xi}_{1},\bsy{\xi}_{2},\nu)\right]\nonumber\\
&& \times \exp \left[-i Kz_{c}\
f_{oo}(\bsy{\xi}_{1},\bsy{\xi}_{2},\nu)\right],
\end{eqnarray}
where $N$ is a normalization constant and
\begin{eqnarray}
\label{foo}
f_{oo}(\bsy{\xi}_{1},\bsy{\xi}_{2},\nu)&=&\mu_{oo}+
\frac{\alpha_{p}}{2}\left[(1+\nu)\xi_{1x}+(1-\nu)\xi_{2x}\right]\nonumber\\
&&-\frac{1}{4\bar{n}_{o}} \Big{\{} (1+\nu) \xi_{1}^{2} + (1-\nu)
\xi_{2}^{2}\nonumber\\
&&-\frac{b}{2}\left[(1+\nu) \xi_{1x}+(1-\nu) 
\xi_{2x}\right]^{2}\nonumber\\
&&-\frac{g}{2}\left[(1+\nu) \xi_{1y}+(1-\nu) 
\xi_{2y}\right]^{2}\Big{\}}.
\end{eqnarray}

\begin{figure}
\centerline{\includegraphics{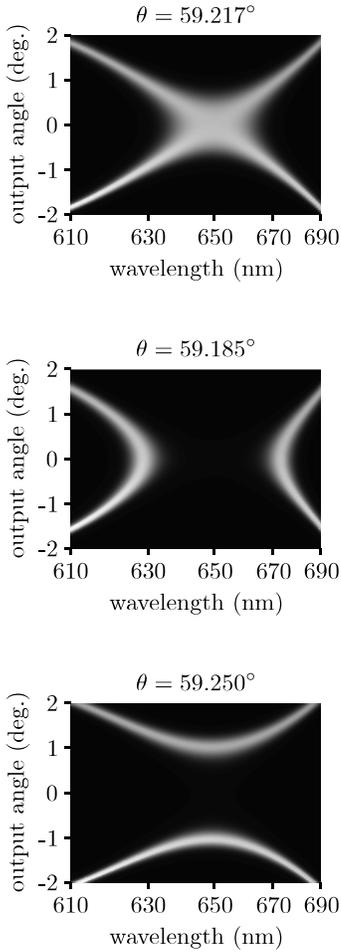}}
\caption{\label{spectrum} Output angle in the $xz$ plane as a 
function of the down-converted wavelength for a 15mm-long Lithium Iodate 
crystal pumped by a 325nm laser beam, for different values of the 
phase matching angle $\theta$. These results were obtained 
from Eqs. (\ref{phiang}) and (\ref{foo}) integrating in 
$\bsy{\xi}_{2}$ with $\xi_{1y}=0$. The (nonlinear) horizontal scale 
was adjusted to allow a direct comparison with experimental data 
reported by Bogdanov \textit{et al.} \cite{bogdanov06}.}
\end{figure}

The single count rate as a function of $\bsy{\xi}_{1}$
($\bsy{\xi}_{2}$) and $\nu$ can be obtained from Eq.  (\ref{phiang})
by integration in $\bsy{\xi}_{2}$ ($\bsy{\xi}_{1}$).  To show that
Eqs.  (\ref{phiang}) and (\ref{foo}) describe the correct dependence
of $\Phi$ on frequency and output angles, the output angle $\xi_{x}$
is plotted as a function of the down-converted wavelength in Fig.
\ref{spectrum} for a 15mm-long Lithium Iodate crystal pumped by a 325nm
laser beam for three different values of the phase matching angle
$\theta$.  Results for $\theta=59.217^{\circ}$ (collinear) and
$\theta=59.185^{\circ}$ are in good agreement with experimental
data reported by Bogdanov \textit{et al.} \cite{bogdanov06} for
$\theta=59.22^{\circ}$ and $\theta=58.97^{\circ}$, respectively.  The
difference of $0.22^{\circ}$ in the noncollinear phase matching angle
is probably due to the fact that the noncollinear condition is
obtained by tilting the crystal.

\begin{figure}
\centerline{\includegraphics{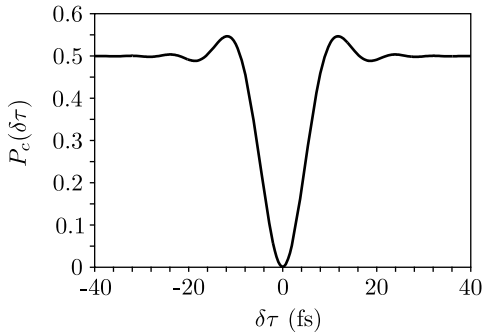}}
\caption{\label{fighom1}Hong-Ou-Mandel dip for a 1mm-long BBO crystal
cut for type I phase matching, with $\theta=34^{\circ}$ pumped by a 351nm
laser, calculated with Eqs.  (\ref{phiang}) and (\ref{hom1}). It is 
assumed that no interference filters are used.}
\end{figure}

The two-photon frequency spectral content of $\Phi_{oo}$ for a given
pair of selected output angles $\bsy{\xi}^{o}_{1}$ and
$\bsy{\xi}^{o}_{2}$ can also be derived from Eq.  (\ref{phiang}).
Experimentally, this spectrum can be directly measured \cite{baek08}, or
inferred from the Hong-Ou-Mandel dip when the measurement is taken
with broadband filters.  The expected profile of the coincidence
detection probability as a function of the path length difference
$\delta l=c\delta\tau$ in a Hong-Ou-Mandel interferometer is given by
\cite{hom}
\begin{equation}
\label{hom1}
P_{c}(\delta\tau)= \frac{1}{2}\left[1-\frac{\int d\tau\,  
\tilde{\phi}^{*}(\tau)\, \tilde{\phi}(\tau-\delta\tau)}{\int d\tau\,
|\tilde{\phi}|^{2}(\tau)}\right],
\end{equation}
where, for a pump beam with a symmetric transverse profile,  
\begin{equation}
\label{phihom}
\tilde{\phi}(\tau)=\int d\nu\ \Phi_{oo}(\bsy{\xi}^{o}_{1},
\bsy{\xi}^{o}_{2},\nu)\ e^{-iK\nu c\delta\tau}.
\end{equation}
The integrals are taken from $-\infty$ to $\infty$.  Fig.
\ref{fighom1} shows a plot of $P_{c}(\delta\tau)$ calculated with Eqs.
(\ref{phiang}) and (\ref{hom1}), for a BBO crystal with $L_{z}=1$mm in
type I phase matching, pumped by a
351nm laser, without interference filters. The collection angles are 
such that both down-converted beams are centered at 702nm.

\begin{figure}
\centerline{\includegraphics{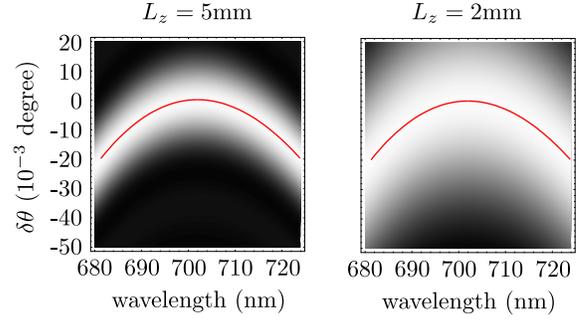}}
\caption{\label{match1}(Color online) Down-converted wavelengths for
collinear ($\bsy{\xi}_{1}=\bsy{\xi}_{2}=0$) type I phase matching as a
function of $\delta \theta$, the deviation from
$\theta_{m}=51.704^{\circ}$ for Lithium Iodate with $L_{z}=5$mm (left)
and $L_{z}=2$mm (right), pumped with $\lambda_{p}=351$nm. The solid 
line corresponds to the condition $\mu_{oo}$=0.}
\end{figure}

\begin{figure}
\centerline{\includegraphics{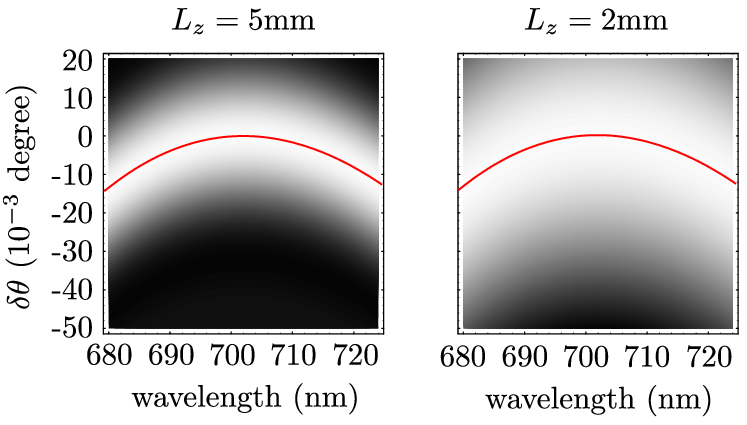}}
\caption{\label{match2}(Color online) Down-converted wavelengths for
collinear ($\bsy{\xi}_{1}=\bsy{\xi}_{2}=0$) type I phase matching as a
function of $\delta \theta$, the deviation from
$\theta_{m}=33.543^{\circ}$ for BBO with $L_{z}=5$mm (left)
and $L_{z}=2$mm (right), pumped with $\lambda_{p}=351$nm. The solid 
line corresponds to the condition $\mu_{oo}$=0.}
\end{figure}

In the collinear detection configuration
($\bsy{\xi}_{1}=\bsy{\xi}_{2}=0$), the spectral profile of the
two-photon state depends on the phase matching angle $\theta$.
Figures \ref{match1} and \ref{match2} show the down-converted
wavelengths for collinear type I phase matching as a function of
$\delta \theta$, the deviation from $\theta_{m}=51.704^{\circ}$ for
Lithium Iodate and from $\theta_{m}=33.543^{\circ}$ for BBO, with
$L_{z}=5$mm and $L_{z}=2$mm, both crystals pumped with
$\lambda_{p}=351$nm.

Another useful representation of $\Phi_{oo}$ is obtained when the
following new coordinates are used:
\begin{subequations}
\label{newvar}
\begin{eqnarray}
\bsy{\xi}_{s}&=& \frac{1}{2}(\bsy{\xi}_{1}+\bsy{\xi}_{2}),\\
\bsy{\xi}_{d}&=& \frac{1}{2}(\bsy{\xi}_{1}-\bsy{\xi}_{2}).
\end{eqnarray}
\end{subequations}
Now, Eq.(\ref{phiang}) is written as
\begin{eqnarray}
\label{phiang2}
\Phi_{oo}& \approx &
N\,G(\nu)\,\tilde{\mathcal{E}}\left(K\bsy{\xi}_{s}+K\nu\bsy{\xi}_{d}\right)\nonumber\\
&& \times \sinc\left[\frac{KL_{z}}{2}\
F_{oo}(\bsy{\xi}_{s},\bsy{\xi}_{d},\nu)\right]\nonumber\\
&& \times \exp \left[-i Kz_{c}\
F_{oo}(\bsy{\xi}_{s},\bsy{\xi}_{d},\nu)\right],
\end{eqnarray}
where $N$ is a normalization constant and
\begin{eqnarray}
\label{Foo}
F_{oo}(\bsy{\xi}_{s},\bsy{\xi}_{d},\nu)&=&\mu_{oo}+
\alpha_{p}(\xi_{sx}+\nu\xi_{dx})\nonumber\\
&& -\frac{1}{2\bar{n}_{o}} [ (1-b) \xi_{sx}^{2} + (1-g)
\xi_{sy}^{2} + \xi_{d}^2\nonumber\\
&& +(2-b)\nu \xi_{sx}\xi_{dx} + (2-g)\nu 
\xi_{sy}\xi_{dy}],\nonumber\\
\end{eqnarray}
where $\nu^2$ was neglected with respect to $\nu$.

Many of the interesting features of $\Phi_{oo}$ are retained when the
anisotropy is neglected in second order,
that is, when the parameters $a$ and $b$ are both approximated by 1.
In this case,
\begin{equation}
\label{Fooapprox}
F_{oo}(\bsy{\xi}_{s},\bsy{\xi}_{d},\nu)\approx\mu_{oo}+
\alpha_{p}(\xi_{sx}+\nu\xi_{dx}) - \frac{1}{2\bar{n}_{o}}
(\xi_{d}^2 +\nu \bsy{\xi}_{s}\cdot\bsy{\xi}_{d}).
\end{equation}
In the frequency degenerate case ($\nu=0$), 
\begin{equation}
\label{Fooapproxdeg}
F_{oo}(\bsy{\xi}_{s},\bsy{\xi}_{d},\nu)\approx\mu_{oo}+
\alpha_{p}\xi_{sx} - \frac{\xi_{d}^2}{2\bar{n}_{o}}.
\end{equation}

Let us analyze the behavior of Eqs. (\ref{phiang2}) and
(\ref{Fooapproxdeg}) in two different measurement schemes: (a) When
the detectors are scanned in opposite directions in the far field, so
that $\bsy{\xi}_{s}=0$.  The probability of coincidence detection is
proportional to
\begin{equation}
\label{cone1}
|\Phi_{oo}|_{d}^{2} \approx\left|N\,G(\nu)\, \tilde{\mathcal{E}}(0)\
\sinc\left[\frac{KL_{z}}{4\bar{n}_{o}}(R^{2} 
-\xi_{d}^{2})\right]\right|^{2}.
\end{equation}
Eq. (\ref{cone1}) describes a circular profile in the variable
$\xi_{d}$, with a radius $R=\sqrt{2\bar{n}_{o}\mu_{oo}}$ and
a half-width $\delta R=\sqrt{R^{2}+4 \pi\bar{n}_{o}/KL_{z}}-R $.  Under
these detection conditions, the coincidence transverse profile maps
the well-known down-converted light cones.

\begin{figure}
\centerline{\includegraphics{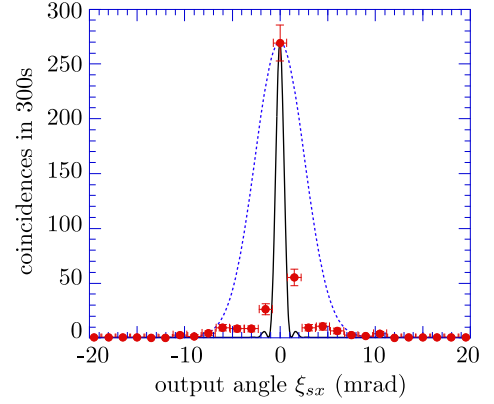}}
\caption{\label{clip1}(Color online) Comparison between experimental
results ($\bullet$) and the prediction of Eq.  (\ref{transf}) (solid
line) for a BBO crystal with $L_{z}=5$mm in collinear type I phase
matching, pumped by a 405nm laser with $w_{0}\approx 25\mu$m.  The two
detectors are scanned in the same sense, along the $x$ direction. The 
dashed line shows a gaussian profile corresponding to the pump laser 
beam.}
\end{figure}

\begin{figure}
\centerline{\includegraphics{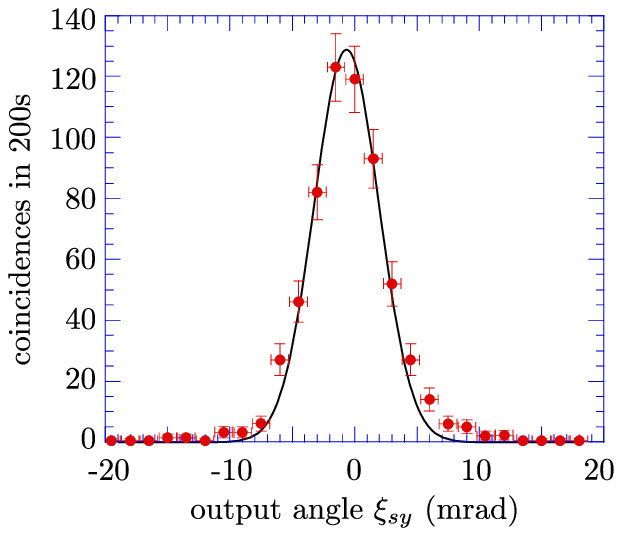}}
\caption{\label{clip2}(Color online) Comparison between experimental
results ($\bullet$) and the prediction of Eq.  (\ref{transf}) (solid
line) for a BBO crystal with $L_{z}=5$mm in collinear type I phase
matching, pumped by a 405nm laser with $w_{0}\approx 25\mu$m.  The two
detectors are scanned in the same sense, along the $y$ direction. The 
Gaussian predicted by Eq.(\ref{transf}) coincides with the expected 
profile of the pump beam.}
\end{figure}

(b) When the detectors are scanned in the same direction in the far
field, so that
$|\bsy{\xi}_{d}|=R$. This condition means
\begin{equation}
\label{transf}
|\Phi_{oo}|_{s}^{2}\approx\left|N\,G(\nu)\,
\tilde{\mathcal{E}}(K \bsy{\xi}_{s})\ 
\sinc\left(\frac{1}{2}KL_{z}\alpha_{p}\xi_{sx}\right)\right|^{2}.
\end{equation}
Eq.  (\ref{transf}) illustrates the possibility of transferring the
angular spectrum $\tilde{\mathcal{E}}(\brm{q})$ from the pump beam to
the down-converted field $\tilde{\mathcal{E}}(K
\bsy{\xi}_{s})=\tilde{\mathcal{E}}(\brm{q}_{1}+\brm{q}_{2})$
\cite{monken98}, and its dependence on the crystal length $L_{z}$.  The
sinc function has a half-width $\delta \xi_{sx}=2\pi/KL_{z}\alpha_{p}$
in the $x$ direction.  Depending on the spatial bandwidth of the pump
beam, this factor may limit the transfer of its angular spectrum to
the two-photon field.  Considering that the pump beam is Gaussian, the
transfer will be satisfactory only when its waist $w_{0}$ is much
larger than $L_{z}\alpha_{p}/2\pi$.  For a BBO of $L_{z}\sim$ 1mm,
$L_{z}\alpha_{p}/2\pi\sim 10\mu$m.  This ``clipping'' in the angular
spectrum transfer, caused by the walk-off term $\alpha_{p}\xi_{sx}$,
has consequences in the anisotropy of entanglement in spatial
variables of the two-photon state \cite{fedorov07}.  Figs.
\ref{clip1} and \ref{clip2} show a comparison between experimental
results obtained in our labs and the predictions of Eq.
(\ref{transf}) for a BBO crystal with $L_{z}=5$mm in collinear type I
phase matching, pumped by a 405nm laser with $w_{0}\approx 25\mu$m.
In Fig.  \ref{clip1}, the two detectors are scanned in the same sense,
along the $x$ direction.  The effect of the walk-off term is evident.
The dashed line shows the expected angular profile of the pump beam.
In Fig.  \ref{clip2}, the two detectors are scanned in the same sense,
along the $y$ direction.  In both cases, interference filters of
FWHM=10nm were used in front of the detectors.

\section{Type II phase matching}

In type II phase matching, the calculation is similar, except for the
fact that one of the down-converted beams has extraordinary
polarization ($e\rightarrow oe$ or $e\rightarrow eo$), that is,
\begin{eqnarray}
\label{psiII}
\ket{\Psi}&=&\int\!d\nu \int\!d\brm{q}_{1}
\int\!d\brm{q}_{2}\
[\Phi_{oe}\ket{\brm{q}_{1},\nu;o}
\ket{\brm{q}_{2},-\nu;e}+\nonumber \\
&& \Phi_{eo}\ket{\brm{q}_{1},\nu;e}
\ket{\brm{q}_{2},-\nu;o}].
\end{eqnarray}

Now, the $k_{z}$ component of the $e$-polarized down-converted wave
has to be written according to expression
(\ref{eq:aprox_paraxial_kz}), that is, for the case $e\rightarrow oe$,
\begin{equation}
k_{pz}\approx \varkappa_{p}-\alpha_{p}
q_{px}-\frac{1}{2\varkappa_{p}}(\beta_{p}
q_{px}^{2}+\gamma_{p} q_{py}^{2}),
\end{equation}
\begin{equation}
k_{1z}\approx k_{1}-\frac{q_{1}^{2}}{2k_{1}},
\end{equation}
\begin{equation}
k_{2z}\approx \varkappa_{2}-\bar{\alpha}
q_{2x}-\frac{1}{2\varkappa_{2}}(\bar{\beta}
q_{2x}^{2}+\bar{\gamma} q_{2y}^{2}),
\end{equation}
where $k_{1}=\bar{n}_{o}(1+a\nu)(1+\nu)\omega_{p}/2c$ and
$\varkappa_{2}=\bar{\eta}(1-a'\nu)(1-\nu)\omega_{p}/2c$. The  factor 
$a'$ is defined analogous to Eq. (\ref{defa}), replacing 
$n_{o}$ by $\eta$.
$\bar{\alpha}$, $\bar{\beta}$, $\bar{\gamma}$ and
$\bar{\eta}$ are calculated with $n_{o}$ and $n_{e}$ taken at
$\omega_{p}/2$.

The amplitudes $\Phi_{oe}$ and $\Phi_{eo}$ have expressions analogous 
to Eq. (\ref{ampli2}), with $\Delta_{z}^{oe}$ and $\Delta_{z}^{eo}$ given by
\begin{subequations}
\label{deltaoe}
\begin{eqnarray}
\Delta_{z}^{oe}&=&K\mu_{oe}
+\alpha_{p}q_{1x}+(\alpha_{p}-\bar{\alpha})q_{2x}\nonumber\\
&&+\frac{1}{\bar{n}_{o}K}\Big[\frac{b}{2}(q_{1x}+q_{2x})^{2}+ 
\frac{g}{2}(q_{1y}+q_{2y})^{2}\nonumber\\
&&-\bar{b}q_{2x}^{2}-\bar{g}q_{2y}^{2}-q_{1}^{2}+(q_{1}^{2}-
\bar{b}q_{2x}^{2}-\bar{g}q_{2y}^{2})\nu\Big],\qquad \\
\Delta_{z}^{eo}&=&K\mu_{eo}+
(\alpha_{p}-\bar{\alpha})q_{1x}+\alpha_{p}q_{2x}\nonumber\\
&&+\frac{1}{\bar{n}_{o}K}\Big[\frac{b}{2}(q_{1x}+q_{2x})^{2}+ 
\frac{g}{2}(q_{1y}+q_{2y})^{2}\nonumber\\
&&-\bar{b}q_{1x}^{2}-\bar{g}q_{1y}^{2}-q_{2}^{2}+(q_{2}^{2}-
\bar{b}q_{1x}^{2}-\bar{g}q_{1y}^{2})\nu\Big],
\end{eqnarray}
where
\end{subequations}
\begin{subequations}
\begin{eqnarray}
\mu_{oe}&=&\frac{\bar{n}_{o}+\bar{\eta}}{2}-\eta_{p}+\nu
\frac{\bar{n}_{o}-\bar{\eta}}{2},\\
\mu_{eo}&=&\frac{\bar{n}_{o}+\bar{\eta}}{2}-\eta_{p}-\nu
\frac{\bar{n}_{o}-\bar{\eta}}{2},
\end{eqnarray}
\end{subequations}
$b=\beta_{p}\bar{n}_{o}/\eta_{p}$,
$g=\gamma_{p}\bar{n}_{o}/\eta_{p}$,
$\bar{b}=\bar{\beta}\bar{n}_{o}/\bar{\eta}$,
$\bar{g}=\bar{\gamma}\bar{n}_{o}/\bar{\eta}$, and
$q_{1}^{2}=|\brm{q}_{1}|^{2}$.  In Eqs.
(\ref{deltaoe}), terms in $a$ and $a'$ were neglected
with respect to 1, and $\nu^{2}$ was neglected with respect to $\nu$.
Note that $\mu_{oe}$ and $\mu_{eo}$ have a linear dependence on $\nu$,
differently from $\mu_{oo}$, whose dependence is on $\nu^{2}$.

In terms of the output angles $\bsy{\xi}_{1}$ and $\bsy{\xi}_{2}$, we have
\begin{eqnarray}
\label{amplioe}
\Phi_{oe}& \approx & N\,G(\nu)\,\tilde{\mathcal{E}}\left[\frac{K}{2}(\bsy{\xi}_{1} +
\bsy{\xi}_{2})+\frac{K\nu}{2}(\bsy{\xi}_{1} - \bsy{\xi}_{2})\right]\nonumber\\
&& \times \sinc\left[\frac{KL_{z}}{2}\
f_{oe}(\bsy{\xi}_{1},\bsy{\xi}_{2},\nu)\right]\nonumber\\
&& \times \exp \left[-i Kz_{c}\
f_{oe}(\bsy{\xi}_{1},\bsy{\xi}_{2},\nu)\right],
\end{eqnarray}
where $N$ is a normalization constant and
\begin{eqnarray}
\label{foe}
f_{oe}&=&\mu_{oe}+
\frac{\alpha_{p}}{2}(1+\nu)\xi_{1x}+
\frac{\alpha_{p}-\bar{\alpha}}{2}(1-\nu)\xi_{2x}\nonumber\\
&&-\frac{1}{4\bar{n}_{o}} \Big{\{} (1+\nu) \xi_{1}^{2} + (1-\nu)(
\bar{b}\,\xi_{2x}^{2}+ \bar{g}\, \xi_{2y}^{2})\nonumber\\
&&-\frac{b}{2}\left[(1+\nu) \xi_{1x}+(1-\nu) 
\xi_{2x}\right]^{2}\nonumber\\
&&-\frac{g}{2}\left[(1+\nu) \xi_{1y}+(1-\nu) 
\xi_{2y}\right]^{2}\Big{\}}.
\end{eqnarray}
$\Phi_{eo}$ has an expression similar to (\ref{amplioe}), with
\begin{eqnarray}
\label{feo}
f_{eo}&=&\mu_{eo}+
\frac{\alpha_{p}-\bar{\alpha}}{2}(1+\nu)\xi_{1x}
+\frac{\alpha_{p}}{2}(1-\nu)\xi_{2x}\nonumber\\
&&-\frac{1}{4\bar{n}_{o}} \Big{\{} (1+\nu)( \bar{b}\,\xi_{1x}^{2}+
\bar{g}\, \xi_{1y}^{2}) + (1-\nu)\xi_{2}^{2}\nonumber\\
&&-\frac{b}{2}\left[(1+\nu) \xi_{1x}+(1-\nu) 
\xi_{2x}\right]^{2}\nonumber\\
&&-\frac{g}{2}\left[(1+\nu) \xi_{1y}+(1-\nu) 
\xi_{2y}\right]^{2}\Big{\}}.
\end{eqnarray}

\begin{figure}[h]
\centerline{\includegraphics{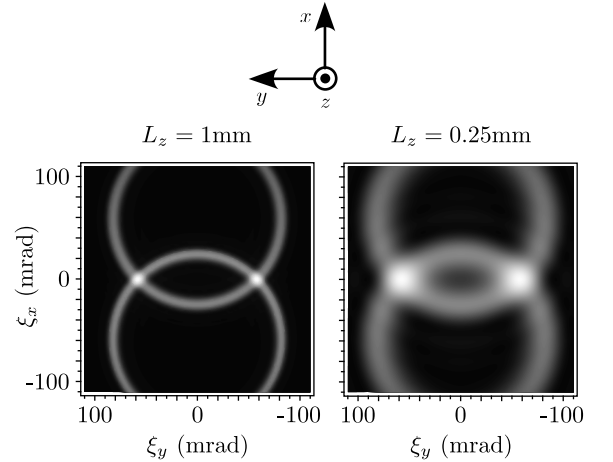}}
\caption{\label{figconeII}Density plot of single counts obtained from Eq.
(\ref{cone2}) integrated in one of the output angles and in $\nu$ over
a bandwidth of 0.025 (which corresponds to a 10nm filter) for
a BBO crystal pumped by a 407nm laser, cut for $\theta=42.5^{\circ}$,
with $L_{z}=1$mm (left) and $L_{z}=0.25$mm (right).}
\end{figure}

Fig.  \ref{figconeII} shows density plots of single counts obtained
from Eqs.  (\ref{amplioe}-\ref{feo}) integrated in one of the output
angles and in $\nu$ over a bandwidth of 0.025 (corresponding to a 10nm
filter) for a BBO crystal pumped by a 407nm laser, cut for
$\theta=42.5^{\circ}$, with $L_{z}=1$mm (left) and $L_{z}=0.25$mm
(right).  Both plots are in good agreement with experimental data
reported by Lee \textit{et al.} \cite{vanexter}.

\begin{figure}
\centerline{\includegraphics{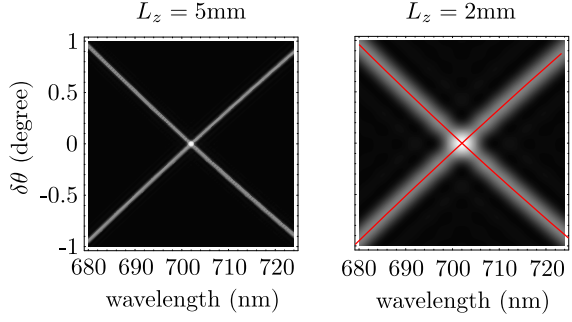}}
\caption{\label{match3} (Color online) Down-converted wavelengths for
collinear ($\bsy{\xi}_{1}=\bsy{\xi}_{2}=0$) type II phase matching as a
function of $\delta \theta$, the deviation from
$\theta_{m}=49.223^{\circ}$ for BBO with $L_{z}=5$mm (left)
and $L_{z}=2$mm (right), pumped with $\lambda_{p}=351$nm. The solid 
lines correspond to the conditions $\mu_{oe}$=0 and $\mu_{eo}$=0.}
\end{figure}

The linear dependence of $\mu_{oe}$ and $\mu_{eo}$ with $\nu$ leads to
a frequency spectrum of the two-photon state quite different from the
type I case.  Figure \ref{match3} shows the down-converted wavelengths
for collinear ($\bsy{\xi}_{1}=\bsy{\xi}_{2}=0$) type II phase matching
as a function of $\delta \theta$, the deviation from
$\theta_{m}=49.223^{\circ}$ for BBO with $L_{z}=5$mm and $L_{z}=2$mm,
pumped with $\lambda_{p}=351$nm. Both plots were obtained from Eqs 
(\ref{amplioe}-\ref{feo}).

\begin{figure}
\centerline{\includegraphics{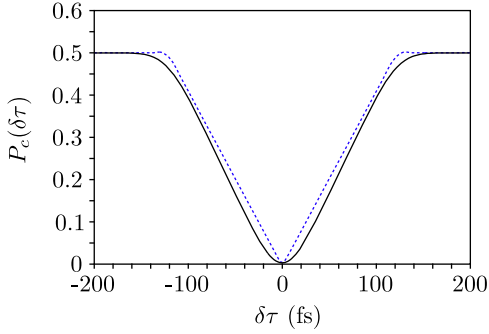}}
\caption{\label{fighom2}(Color online) Hong-Ou-Mandel dip for a
1mm-long BBO crystal cut for type II phase matching, with
$\theta=48.34^{\circ}$ pumped by a 351nm laser, calculated with Eqs.
(\ref{hom1}) and (\ref{amplioe}).  The solid line corresponds to the
case in which interference filters of 20nm bandwidth are used.  The
dashed line corresponds to the case in which no interference filters
are used.}
\end{figure}

Fig.  \ref{fighom2} shows the Hong-Ou-Mandel dip for a
1mm-long BBO crystal cut for \textit{beamlike} type II phase matching
($\theta=48.34^{\circ}$), pumped by a 351nm laser.  The solid line
corresponds to the case in which interference filters of 20nm
bandwidth are used, in good agreement with the data reported by Kim  
\cite{kim}.  The dashed line corresponds to the case in which
no interference filters are used.  Notice the difference between Figs.
\ref{fighom1} and \ref{fighom2}. In type II phase matching, the 
Hong-Ou-Mandel dip is much larger than in type I, indicating a 
narrower frequency spectrum in type II.

It is also possible to write $\Phi_{oe}$ and $\Phi_{eo}$ in terms of
the variables $\bsy{\xi}_{s}$ and $\bsy{\xi}_{d}$ defined in Eq.
\ref{newvar}, but the exact expressions are too long to be of
practical use.  However, when $b$, $\bar{b}$, $g$ and $\bar{g}$ are
all approximated by 1, we arrive at the useful expressions
\begin{subequations}
\label{Foe}
\begin{eqnarray}
F_{oe}&=&\mu_{oe}+
\left (\alpha_{p}-\frac{1+\nu}{2}\bar{\alpha}\right )\xi_{sx}\nonumber\\
&&+\left (\nu \alpha_{p}+\frac{1-\nu}{2}\bar{\alpha}\right )\xi_{dx} -
\frac{\xi_{d}^{2}}{2\bar{n}_{o}},\\
F_{eo}&=&\mu_{eo}+
\left (\alpha_{p}-\frac{1-\nu}{2}\bar{\alpha}\right )\xi_{sx}\nonumber\\
&&+\left (\nu \alpha_{p}-\frac{1+\nu}{2}\bar{\alpha}\right)\xi_{dx} -
\frac{\xi_{d}^{2}}{2\bar{n}_{o}}.
\end{eqnarray}
\end{subequations}
When the detectors are scanned in opposite directions in the far
field, so that $\bsy{\xi}_{s}=0$,  the probability of coincidence
 detection $P_{d}$ is proportional to 
$\int d\nu\, (|\Phi_{oe}|_{d}^{2}+|\Phi_{eo}|_{d}^{2})$, where
\begin{subequations}
\label{cone2}
\begin{eqnarray}
|\Phi_{oe}|_{d}^{2}&\approx&  |N\,G(\nu)\, 
\tilde{\mathcal{E}}(0)|^{2}\nonumber\\
&&\times\,\sinc^{2}\Big [\frac{KL}{4\bar{n}_{o}}(R_{oe}^{2}-|\bsy{\xi}_{d}-
\bsy{c}_{oe}|^{2})\Big ],\\
|\Phi_{eo}|_{d}^{2}&\approx&  |N\,G(\nu)\, 
\tilde{\mathcal{E}}(0)|^{2}\nonumber\\
&&\times\,\sinc^{2}\Big [\frac{KL}{4\bar{n}_{o}}(R_{eo}^{2}-|\bsy{\xi}_{d}-
\bsy{c}_{eo}|^{2})\Big ],\ \ \  
\end{eqnarray}
\end{subequations}
\begin{subequations}
\begin{eqnarray}
R_{oe}^{2}&=&2\bar{n}_{o} \mu_{oe} + {\bar{n}_{o}^{2} 
\bar{\alpha}^{2}} \left [\frac{1}{4} - \nu\left(\frac{\alpha_{p}}{\bar{\alpha}}
-\frac{1}{2}\right)\right],\\
R_{eo}^{2}&=&2\bar{n}_{o} \mu_{eo} + {\bar{n}_{o}^{2} 
\bar{\alpha}^{2}} \left [\frac{1}{4} + \nu\left(\frac{\alpha_{p}}{\bar{\alpha}}
-\frac{1}{2}\right)\right],
\end{eqnarray}
\end{subequations}
\begin{subequations}
\begin{eqnarray}
\bsy{c}_{oe}&=&\bar{n}_{o}\bar{\alpha}\left [\frac{1}{2} +
\nu\left(\frac{\alpha_{p}}{\bar{\alpha}} -\frac{1}{2}\right)\right]
\brm{\hat{x}},\\
\bsy{c}_{eo}&=&\bar{n}_{o}\bar{\alpha}\left [-\frac{1}{2} +
\nu\left(\frac{\alpha_{p}}{\bar{\alpha}} -\frac{1}{2}\right)\right]
\brm{\hat{x}},
\end{eqnarray}
\end{subequations}
and $\brm{\hat{x}}$ is the unit vector in the $x$ direction.

Expressions (\ref{cone2}) describe two circular profiles, centered at
$\bsy{\xi}_{d}=\bsy{c}_{oe}$ and $\bsy{\xi}_{d}=\bsy{c}_{eo}$, with
radii $R_{oe}$ and $R_{eo}$, respectively.  Collinear down-conversion
occurs when $R_{oe}=c_{oe}$ and $R_{eo}=c_{eo}$.  The so-called
\textit{beamlike} type II down-conversion occurs in the degenerate
case ($\nu=0$) when $\theta$ is such that $R_{oe}=R_{eo}=0$.

\begin{figure}
\centerline{\includegraphics{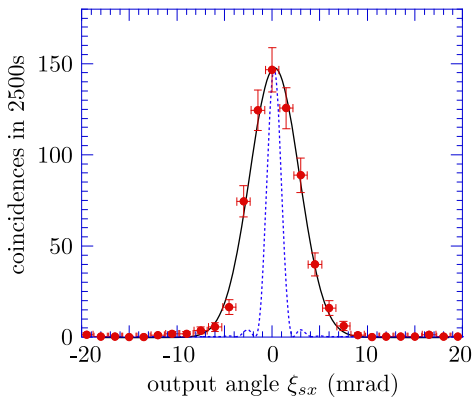}}
\caption{\label{clip3} (Color online)Comparison between experimental
results ($\bullet$) and the prediction of Eq.  (\ref{transf2}) (solid
line) for a BBO crystal with $L_{z}=5$mm in collinear type II phase
matching, pumped by a 405nm laser with $w_{0}\approx 25\mu$m.  The two
detectors are scanned in the same sense, along the $x$ direction.  The
Gaussian predicted by Eq.(\ref{transf2}) coincides with the expected
profile of the pump beam.  The dashed line shows the expected curve
for the monochromatic case.}
\end{figure}

\begin{figure}
\centerline{\includegraphics{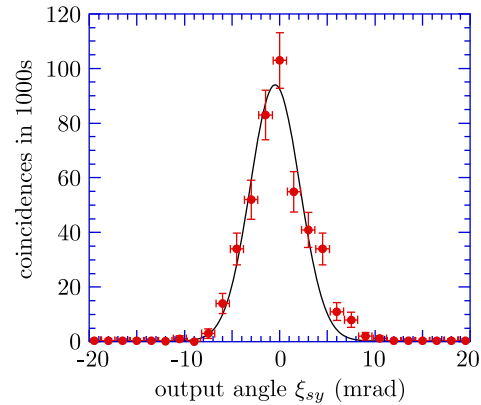}}
\caption{\label{clip4} (Color online)Comparison between experimental
results ($\bullet$) and the prediction of Eq.  (\ref{transf2}) (solid
line) for a BBO crystal with $L_{z}=5$mm in collinear type II phase
matching, pumped by a 405nm laser with $w_{0}\approx 25\mu$m.  The two
detectors are scanned in the same sense, along the $y$ direction. The 
Gaussian predicted by Eq.(\ref{transf2}) coincides with the expected 
profile of the pump beam.}
\end{figure}

Let us analyze the case when the detectors are scanned in the same 
direction in the far field, so that $\bsy{\xi}_{d}=0$. Then, 
\begin{subequations}
\label{transf2}
\begin{eqnarray}
|\Phi_{oe}|_{s}^{2}&\approx& |N\,G(\nu)\, \tilde{\mathcal{E}}(K 
\bsy{\xi}_{s})|^{2}\nonumber\\
&&\times\,\sinc^{2}\Big{\{}\frac{KL_{z}}{2}\Big{[}\mu_{oe}
+ \Big{(}\alpha_{p} - 
\frac{\bar{\alpha}}{2}\Big{)}\xi_{sx}\Big{]}\Big{\}},\ \ \  \\
|\Phi_{eo}|_{s}^{2}&\approx& |N\,G(\nu)\, \tilde{\mathcal{E}}(K 
\bsy{\xi}_{s})|^{2}\nonumber\\
&&\times\,\sinc^{2}\Big{\{}\frac{KL_{z}}{2}\Big{[}\mu_{eo}
+ \Big{(}\alpha_{p} - 
\frac{\bar{\alpha}}{2}\Big{)}\xi_{sx}\Big{]}\Big{\}}.\ \ \
\end{eqnarray}
\end{subequations}
Note that in type II the transfer of angular spectrum described by
Eqs.  (\ref{transf2}) is affected by the detuning $\nu$ through the
parameters $\mu_{oe}$ and $\mu_{eo}$.  The detuning has the effect of
displacing the sinc function by an amount
$(\bar{n}_{o}-\bar{\eta})KL_{z}\nu/4$ in the $x$ axis, in opposite
directions for $|\Phi_{oe}|_{s}^{2}$ and $|\Phi_{eo}|_{s}^{2}$.  In
the limit of long crystals, the sinc$^{2}$ functions are narrow enough
to allow us to write the coincidence detection probability as
\begin{eqnarray}
P_{s}&=&\int d\nu\, (|\Phi_{oe}|_{s}^{2}
+|\Phi_{eo}|_{s}^{2})\nonumber\\
&\approx&|N\, \tilde{\mathcal{E}}(K
\bsy{\xi}_{s})|^{2}\ \left[ G^{2}\Big{(}-\frac{\xi_{sx}}{m}\Big{)}+
 G^{2}\Big{(}\frac{\xi_{sx}}{m}\Big{)}\right],
\end{eqnarray}
where $m=(2\alpha_{p} - \bar{\alpha})/(\bar{n}_{o} - \bar{\eta})$.  In
type II phase matching, due to the linear dependence of $\mu_{oe}$ and
$\mu_{eo}$ on $\nu$, the angular spectrum of the pump field may be
transferred to the two-photon state if the bandwidth of the detection
filters is broad enough, even for long crystals.  Figs.  \ref{clip3}
and \ref{clip4} show comparison of the predictions of Eq.
(\ref{transf2}) with experimental data for a 5mm-long BBO crystal
pumped by a 405nm laser with $w_{0}\approx 25\mu$m and detection
filters with a bandwidth of $10$nm.  In both figures, there is a good
agreement between experimental data and the expected Gaussian profile
of the pump beam.  The experimental conditions are the same of Figs.
\ref{clip1} and \ref{clip2}, except for the phase matching type.
Notice the difference between the two cases when the detectors are
scanned in the $x$ direction (Figs.  \ref{clip1} and \ref{clip3}).  In
the case of Fig.  \ref{clip3}, the 10nm filter allows a complete
transfer of the pump beam profile to the coincidence detection.  The
dashed line shows the expected curve for the monochromatic case, that
is, $G(\nu)\approx\delta(\nu)$ (zero bandwidth filter).

\section{Summary and conclusion}

We have discussed in detail the two-photon state generated by
spontaneous parametric down-conversion in bulk crystals, taking into
account the effect of crystal anisotropy, in both in type I and type
II phase matching.  Our discussion was based on the perturbative
approach introduced by Hong and Mandel, in the context of Fourier
Optics, in which the fields are treated by means of their plane-wave
expansion.  The frequency spectrum of the down-converted fields was
also considered, in the approximation of small detuning and
monochromatic pump beams.  Extension of the theory to cover pulsed
pump beams seems straightforward.  Several approximations were made in
order to provide simple expressions that still exhibit the main
features of the frequency and spatial spectral properties of the
two-photon states.  These expressions were shown to be in good
agreement with experimental data for some selected situations.  In
particular, the spectral content of the two-photon state and the
transfer of angular spectrum from the pump beam to the two-photon
state were analyzed, and a significant differences of these features in
type I and type II phase matching were discussed.  The results
presented here are a contribution to the Fourier optics of two-photon
states and may be helpful to improve the understanding and further
development of entangled state sources for quantum information and
quantum communication.
\begin{acknowledgments}
This work was supported by the
Brazilian funding agencies CNPq, CAPES and FAPEMIG.  
\end{acknowledgments}

\end{document}